\documentclass[aps,showpacs,prd,twocolumn]{revtex4}%
\usepackage{graphicx}
\usepackage{amssymb}
\usepackage{amsmath}
\usepackage{float}
\usepackage{accents}
\usepackage{graphicx}
\usepackage{color}
\usepackage{amsfonts}%
\providecommand{\U}[1]{\protect\rule{.1in}{.1in}}
\begin{document}
\title{Deformed self-dual magnetic monopoles}
\author{D. Bazeia$^{1,2}$, R. Casana$^{3}$, M. M. Ferreira Jr.$^{3}$, E. da
Hora$^{3,4}$ and L. Losano$^{1,2}$}
\affiliation{$^{1}$Departamento de F\'{\i}sica, Universidade Federal da Para\'{\i}ba,
58051-900, Jo\~{a}o Pessoa, Para\'{\i}ba, Brazil}
\affiliation{$^{2}$Departamento de F\'{\i}sica, Universidade Federal de Campina Grande,
58109-970, Campina Grande, Brazil}
\affiliation{$^{3}$Departamento de F\'{\i}sica, Universidade Federal do Maranh\~{a}o,
65085-580, S\~{a}o Lu\'{\i}s, Maranh\~{a}o, Brazil}
\affiliation{$^{4}$Coordenadoria do Curso Interdisciplinar em Ci\^{e}ncia e Tecnologia,
Universidade Federal do Maranh\~{a}o, 65080-805, S\~{a}o Lu\'{\i}s,
Maranh\~{a}o, Brazil}
\date{\today}

\begin{abstract}
We develop a deformation method for attaining new magnetic monopole analytical
solutions consistent with generalized Yang-Mills-Higgs model introduced
recently. The new solutions fulfill the usual radially symmetric ansatz and
the boundary conditions suitable to assure finite energy configurations. We
verify our prescription by studying some particular cases involving both
exactly and partially analytical initial configurations whose deformation
leads to new analytic BPS monopoles. The results show consistency among the
models, the deformation procedure and the profile of the new solutions.

\end{abstract}

\maketitle

\section{Introduction}

\label{Intro}

In the context of classical field theories, configurations possessing
nontrivial topology are usually described as static solutions to some
nonlinear models \cite{n5}. In particular, these models usually allow for the
spontaneous symmetry breaking mechanism, since ordinary topological defects
are known to be formed during symmetry breaking phase transitions. Beyond
that, in very special cases, topologically nontrivial structures can be
obtained by solving a given set of first-order differential equations
\cite{n4}. In addition, one also verifies that the solutions obtained this way
possess the minimum energy possible, since they saturate a given lower bound
for the total energy.

In this sense, the simplest topological defect is the static kink \cite{n0}
appearing within a classical model containing one single real scalar field.
Also, regarding higher dimensional scenarios, the ordinary vortex \cite{n1}
arises within a planar Abelian-Higgs model, whilst the magnetic monopole
\cite{n3} stands for the topological profile coming from a (1+3)-dimensional
non-Abelian-Higgs theory.

In addition, during the last years, topological solutions arising within
nonstandard field models have been intensively studied, such models being
endowed by noncanonical kinetic terms which change the overall dynamics in a
nonusual way. The interesting point is that these theories engender
topological configurations even in the absence of symmetry breaking potentials
for the matter self-interaction. It is worthwhile to point out that the idea
regarding generalized dynamics arises in a rather natural way in the context
of the string theories. Furthermore, these new results have been applied to
many physical investigations, including the ones regarding the accelerated
inflationary phase of the universe \cite{n8}, strong gravitational waves
\cite{sgw}, tachyon matter \cite{tm}, dark matter \cite{dm}, and others
\cite{o}.

Recently, some of us have investigated the way the noncanonical scenarios
engender self-duality \cite{o1}. The overall conclusion is that, in general,
the new self-dual solutions behave in the same way their standard partners do.
However, within some particular cases, unusual kinetic terms also change the
shape of the engendered profiles by inducing variations on the defect
amplitude and characteristic length. Many additional properties of such
theories and their solutions can be found in Ref. \cite{o2}. In particular, it
was also verified the way the generalized theories mimic the standard results,
the so-called twinlike models \cite{twin}.

On the other hand, some years ago, some of us have introduced a particular
prescription, named the deformation method \cite{d1}, which allows for the
calculation of new models starting from well-established ones. The overall
prescription relies on an invertible and differentiable deformation function,
to be chosen conveniently. The method was initially proposed for the study of
(1+1)-dimensional theories containing scalar fields only. In this sense,
deformed solutions were already investigated within polynomial \cite{d2},
sine-Gordon and multi-sine-Gordon scenarios \cite{d4}. Besides, an orbit-based
extension of such prescription was applied to models involving two interacting
scalar fields \cite{d5}. More recently, similar calculations regarding the
static domain walls arising in a noncanonical Abelian-Chern-Simons-Higgs model
were also performed \cite{d6}.

In this letter, we go further by introducing a deformation prescription
consistent with the generalized non-Abelian-Higgs model firstly introduced in
\cite{pau}. In order to present our results, this paper is organized as
follows. In Sec. II, we review the way the nonstandard Yang-Mills-Higgs
theory, that we consider as our starting-point, engenders self-duality. The
non-Abelian fields are supposed to be described by the usual spherically
symmetric ansatz, the corresponding solutions standing for BPS\ magnetic
monopoles possessing finite energy. In the sequel, in Sec. III, we attain our
main goal by introducing a deformation prescription consistent with this
non-Abelian theory. Further, in Sec. IV, we verify our construction by
studying some particular examples. Here, it is worthwhile to say that the
prescription we have introduced works very well for both totally and partially
analytical scenarios, the deformed configurations being well-behaved in all
relevant sectors. Finally, in Sec. V, we present our concluding remarks and
perspectives regarding future investigations.


\section{\textbf{The basic model}}

\label{general0}

We begin reviewing the investigation performed in Ref. \cite{pau}, whose
starting-point is the (1+3)-dimensional Lagrangian density%
\begin{equation}
\mathcal{L}=-\frac{G\left(  \phi^{a}\phi^{a}\right)  }{4}F_{\mu\nu}^{b}%
F^{\mu\nu,b}+\frac{M\left(  \phi^{a}\phi^{a}\right)  }{2}D_{\mu}\phi^{b}%
D^{\mu}\phi^{b}\text{,}\label{1}%
\end{equation}
where $F_{\mu\nu}^{a}=\partial_{\mu}A_{\nu}^{a}-\partial_{\nu}A_{\mu}%
^{a}+e\epsilon^{abc}A_{\mu}^{b}A_{\nu}^{c}$ stands for the non-Abelian field
strength tensor, $G\left(  \phi^{a}\phi^{a}\right)  $ and $M\left(  \phi
^{a}\phi^{a}\right)  $ are arbitrary positive functions which generalize the
overall dynamics of the model. Also, $D_{\mu}\phi^{a}=\partial_{\mu}\phi
^{a}+e\epsilon^{abc}A_{\mu}^{b}\phi^{c}$\ is the non-Abelian covariant
derivative and $\epsilon^{abc}$\ is the totally antisymmetric Levi-Civita
symbol. The Lagrangian density above can be seen as the low energy limit of a
supersymmetric field theory, involving non Abelian fields coupled to gravity
\cite{NPB}. It can be also considered as an effective field model describing
the dynamics of non Abelian fields in a chromoelectric media whose properties
are defined by the functions $G\left(  \phi^{a}\phi^{a}\right)  $\ and
$M\left(  \phi^{a}\phi^{a}\right)  $\ \cite{pau}\textbf{. }Along the paper, we
use standard conventions, including the plus-minus signature for the Minkowski
space-time. For simplicity, along this paper, all fields, coordinates and
parameters are considered to be dimensionless, and we fix $e=1$.

This work is devoted to the study of static uncharged (the temporal gauge,
$A_{0}^{a}=0$, satisfies trivially the Gauss law of the non-Abelian model)
configurations with spherically symmetric solutions arising from (\ref{1}),
which can be implemented via the standard ansatz%
\begin{equation}
\phi^{a}=x^{a}\frac{H\left(  r\right)  }{r}\text{,} \label{a}%
\end{equation}%
\begin{equation}
A_{i}^{a}=\epsilon_{iak}x_{k}\frac{W\left(  r\right)  -1}{r^{2}}\text{,}
\label{b}%
\end{equation}
where $r^{2}=x_{a}x^{a}$. Consequently, the profile functions $H\left(
r\right)  $ and $W\left(  r\right)  $ are supposed to obey the following
boundary conditions:%
\begin{equation}
H\left(  0\right)  =0\text{ \ \ and \ \ }W\left(  0\right)  =1\text{,}
\label{cc1}%
\end{equation}%
\begin{equation}
H\left(  \infty\right)  =\mp1\text{ \ \ and \ \ }W\left(  \infty\right)
=0\text{,} \label{cc2}%
\end{equation}
guaranteeing the spontaneous breaking of the SO(3) symmetry inherent to
(\ref{1}). Thus, the functions $H\left(  r\right)  $ and $W\left(  r\right)  $
describe topological solutions possessing finite total energy.

In Ref. \cite{pau}, it was verified that the non-Abelian model (\ref{1}) only
yields self-dual solutions when $G\left(  \phi^{a}\phi^{a}\right)  $ and
$M\left(  \phi^{a}\phi^{a}\right)  $ satisfy the following constraint:%
\begin{equation}
G=\frac{1}{M}\text{.}\label{v}%
\end{equation}
In order to review the way the self-duality happens, we point out that, when
considering (\ref{v}), the static energy density related to (\ref{1}) can be
written in the form (already supposing the temporal gauge)%
\begin{equation}
\varepsilon=\frac{1}{4M}\left(  F_{ik}^{a}\pm\epsilon_{ikj}MD_{j}\phi
^{a}\right)  ^{2}\mp\frac{1}{2}\epsilon_{ikj}F_{ik}^{a}D_{j}\phi^{a}%
\text{,}\label{en}%
\end{equation}
with the Latin letters, $i$, $k$\ and $j,$\ standing for spatial
coordinates.\ The corresponding total energy is minimized by the self-dual
equation%
\begin{equation}
F_{ik}^{a}\pm\epsilon_{ikj}MD_{j}\phi^{a}=0\text{,}\label{nbps}%
\end{equation}
the last term in Eq. (\ref{en}) being the energy density inherent to the
self-dual configurations, i.e.,%
\begin{equation}
\varepsilon_{bps}=\mp\frac{1}{2}\epsilon_{ikj}F_{ik}^{a}D_{j}\phi^{a}%
\text{.}\label{nbpsx}%
\end{equation}

Moreover, given the spherically symmetric ansatz (\ref{a}) and (\ref{b}), the
self-dual equation (\ref{nbps}) provides%
\begin{equation}
\frac{dH}{dr}=\mp\frac{P\left(  r\right)  }{r^{2}}\text{,} \label{bps1}%
\end{equation}%
\begin{equation}
\frac{dW}{dr}=\pm MHW\text{,} \label{bps2}%
\end{equation}
where we have defined the auxiliary function $P\left(  r\right)  $ as%
\begin{equation}
P\left(  r\right)  =\frac{1-W^{2}}{M}\text{.}%
\end{equation}
Thus, the profile functions $H\left(  r\right)  $ and $W\left(  r\right)  $
stand for the solutions of a set of two coupled first-order equations coming
from the minimization of the non-Abelian total energy. Equations (\ref{bps1})
and (\ref{bps2}) are the spherically symmetric BPS ones arising within the
noncanonical Yang-Mills-Higgs scenario (\ref{1}). Once the BPS equations
(\ref{bps1}) and (\ref{bps2}) are considered, the BPS energy density
(\ref{nbpsx}) reduces to%
\begin{equation}
\varepsilon_{bps}=\mp\frac{1}{r^{2}}\frac{d}{dr}\left(  H\left(
1-W^{2}\right)  \right)  \text{,} \label{buc}%
\end{equation}
whilst the total energy is%
\begin{equation}
E_{bps}=4\pi\int r^{2}\varepsilon_{bps}dr=4\pi\text{,} \label{te}%
\end{equation}
whenever the boundary conditions (\ref{cc1}) and (\ref{cc2}) are fulfilled.

In Ref. \cite{pau}, for a particular choice of $M$, some of us have integrated
the first-order equations (\ref{bps1}) and (\ref{bps2}) numerically by means
of the relaxation technique, the resulting solutions being generalized
self-dual magnetic monopoles possessing finite total energy given by Eq.
(\ref{te}). In Ref. \cite{PLB}, one has investigated some effective
non-Abelian models for which the resulting BPS equations were solved
analytically. These analytical profiles behave in the same general way as the
usual ones do, despite one of them has presented a nonstandard ringlike BPS
energy density (which differs from the usual lump-like one).

In the following Section, we go further by introducing a consistent
prescription through which one can always deform a given self-dual monopole
solution into a new one. As we demonstrate, the initial configuration can be
completely analytical (possessing exact solutions for both $H\left(  r\right)
$ and $W\left(  r\right)  $), or only partially analytical (possessing an
exact solution to $H\left(  r\right)  $, but a numerical solution for
$W\left(  r\right)  $).


\section{\textbf{The deformation prescription}}

\label{general}

Here, we develop the deformation prescription for self-dual magnetic monopoles
following the procedure introduced for scalar fields \cite{d1}, also extended
for the Higgs and Abelian gauge fields \cite{d6}. Let us describe the
procedure we will implement to find new monopole solutions. For such purpose,
we firstly suppose a new Lagrangian density mathematically similar to
(\ref{1}), but with new functions $G\rightarrow\mathcal{G}$ and $M\rightarrow
\mathcal{M}$. We still assume that the new scalar and gauge fields are also
described by the spherically symmetric ansatz of eqs. (\ref{a}) and (\ref{b}).
Similarly, the new profile functions $\mathcal{H}\left(  r\right)  $ and
$\mathcal{W}\left(  r\right)  $ obey the same finite energy boundary
conditions pointed in eqs. (\ref{cc1}) and (\ref{cc2}), i.e.,
\begin{equation}
\mathcal{H}\left(  0\right)  =0\text{ \ \ and \ \ }\mathcal{W}\left(
0\right)  =1\text{,} \label{ncc1}%
\end{equation}%
\begin{equation}
\mathcal{H}\left(  \infty\right)  =\mp1\text{ \ \ and \ \ }\mathcal{W}\left(
\infty\right)  =0\text{.} \label{ncc2}%
\end{equation}
Within this scenario, the corresponding BPS equations can be calculated in the
very same way as performed for the initial model (\ref{1}), that is, by
requiring the minimization of the total energy. This leads to the new
self-dual equations
\begin{equation}
\frac{d\mathcal{H}}{dr}=\mp\frac{\mathcal{P}\left(  r\right)  }{r^{2}}\text{,}
\label{nbps1}%
\end{equation}%
\begin{equation}
\frac{d\mathcal{W}}{dr}=\pm\mathcal{MHW}\text{,} \label{nbps2}%
\end{equation}
with $\mathcal{P}\left(  r\right)  $ being given by
\begin{equation}
\mathcal{P}\left(  r\right)  =\frac{1-\mathcal{W}^{2}}{\mathcal{M}}\text{.}
\label{pc}%
\end{equation}
Nevertheless, one also gets that the energy density of the resulting BPS
structures reduces to
\begin{equation}
\mathcal{E}_{bps}=\mp\frac{1}{r^{2}}\frac{d}{dr}\left(  \mathcal{H}\left(
1-\mathcal{W}^{2}\right)  \right)  , \label{nbuc1}%
\end{equation}
so that they possess the same total energy given in Eq. (\ref{te}). Here, we
reinforce that, whereas $M(H)$\ and $\mathcal{M}(\mathcal{H})$\ are not
necessarily equal to each other, the self-dual solutions coming from the two
non-Abelian models are essentially different. In what follows, for simplicity,
we consider only the lower signs in eqs. (\ref{cc2}), (\ref{bps1}),
(\ref{bps2}), (\ref{buc}), (\ref{ncc2}), (\ref{nbps1}), (\ref{nbps2}) and
(\ref{nbuc1}).

We continue our construction by adopting the fundamental relation%
\begin{equation}
H\left(  r\right)  =f\left(  \mathcal{H}\left(  r\right)  \right)  \text{,}
\label{fa}%
\end{equation}
where $f$ stands for an invertible and differentiable deformation function, to
be chosen conveniently. In this case, since $H$ and $f$ are supposed to be
known, the new profile function $\mathcal{H}\left(  r\right)  $ can be
trivially obtained via
\begin{equation}
\mathcal{H}\left(  r\right)  =f^{-1}\left(  H\left(  r\right)  \right)
\text{.} \label{npf}%
\end{equation}
Also, by differentiating (\ref{fa}) with respect to $r$, and using
(\ref{bps1}) and (\ref{nbps1}), the auxiliary functions $P\left(  r\right)  $
and $\mathcal{P}\left(  r\right)  $ obey
\begin{equation}
\mathcal{P}\left(  r\right)  =\frac{P\left(  r\right)  }{f^{\prime}\left(
\mathcal{H}\left(  r\right)  \right)  }\text{,} \label{afr}%
\end{equation}
where $f^{\prime}=df/dH$. Furthermore, by combining (\ref{nbps2}) and
(\ref{pc}), the resulting expression can be integrated to yield%
\begin{equation}
\mathcal{W}\left(  r\right)  =\frac{Ce^{\mathcal{N}\left(  r\right)  }}%
{\sqrt{1+C^{2}e^{2\mathcal{N}\left(  r\right)  }}}\text{,} \label{nw}%
\end{equation}
where $C$ is an integration constant, and the function $\mathcal{N}\left(
r\right)  $\ is%
\begin{equation}
\mathcal{N}\left(  r\right)  =-\int^{r}\frac{\mathcal{H}\left(  r^{\prime
}\right)  }{\mathcal{P}\left(  r^{\prime}\right)  }dr^{\prime}\text{.}
\label{n}%
\end{equation}
From eqs. (\ref{pc}) and (\ref{nw}), we get the constraint (\ref{v}) for the
deformed system%
\begin{equation}
{\mathcal{G}}={\mathcal{M}}^{-1}={\mathcal{P}}(r)\left(  1+C^{2}%
e^{2\mathcal{N}(r)}\right)  \text{.} \label{vd}%
\end{equation}
The basic equations we have to keep in mind are (\ref{npf}), (\ref{afr}),
(\ref{nw}), (\ref{n}) and (\ref{vd}). Here, it is worthwhile to point out that
the only initial data we need to perform our calculation is the analytical
solution for $H\left(  r\right)  $. In this sense, the solution for $W\left(
r\right)  $ can be analytical or even numerical; in both cases, the deformed
scenario will be completely analytical (possessing analytical solutions to
both $\mathcal{H}\left(  r\right)  $ and $\mathcal{W}\left(  r\right)  $).

It is important to clarify that, since the deformed solutions also obey
the\ boundary conditions (\ref{ncc1}) and (\ref{ncc2}), they stand for
nontrivial self-dual magnetic monopoles possessing energy density given by
(\ref{nbuc1}) and finite total energy equal to $4\pi$. Besides that, we also
point out that, in all the new scenarios, the generalization function
$\mathcal{M}$ (or $M$) is positive,\ as required for the non-Abelian model
(\ref{1}) to attain a positive energy density \cite{pau}.

In the next Section, we present our results, including the deformation of a
partially analytical configuration into a completely analytical one.


\section{\textbf{Deformed BPS monopoles}}

\label{general copy(2)}

In order to present our algorithm in an illustrative way, we first apply the
deformation procedure in a completely analytical scenario. The first situation
we address is the deformation of the usual 't Hooft-Polyakov monopole. Indeed,
we have verified that such deformation is possible and that the resulting
configuration has already been obtained in a previous work; see eqs. (16) and
(17) in Ref. \cite{PLB}. Thus, in order to explain the way it happens, we
consider as the starting-point the standard monopole solution:
\begin{equation}
H_{_{tHP}}\left(  r\right)  =\frac{1}{\tanh\left(  r\right)  }-\frac{1}%
{r}\text{,} \label{u1aa}%
\end{equation}%
\begin{equation}
W_{_{tHP}}\left(  r\right)  =\frac{r}{\sinh\left(  r\right)  }\text{.}
\label{u1b}%
\end{equation}
In this case, one takes the function $f$\ as the simplest choice: \
\begin{equation}
f\left(  \mathcal{H}\right)  =\mathcal{H}\left(  r\right)  \text{,}%
\end{equation}
which means that the deformed scenario is described by
\begin{equation}
\mathcal{H}\left(  r\right)  =H_{_{tHP}}\left(  r\right)  \text{.}
\label{dsh1}%
\end{equation}
In this case, despite the usual solution for the Higgs sector, the integration
constant appearing in (\ref{nw}) allows to generalize\ the corresponding
solution for the gauge field, leading to
\begin{equation}
\mathcal{W}\left(  r\right)  =\frac{r}{\sqrt{3w_{0}\sinh^{2}\left(  r\right)
-\left(  3w_{0}-1\right)  r^{2}}}\text{,} \label{dsw1}%
\end{equation}
where $w_{0}>0$ is related to the aforecited integration constant. The
auxiliary functions $\mathcal{P}\left(  r\right)  $ and $\mathcal{N}\left(
r\right)  $, can be obtained via eqs. (\ref{afr}) and (\ref{n}), yielding
\begin{align}
\mathcal{P}\left(  r\right)   &  =\frac{\sinh^{2}\left(  r\right)  -r^{2}%
}{\sin^{2}\left(  r\right)  }\text{,}\label{ax1}\\
\mathcal{N}\left(  r\right)   &  =\ln\left[  \frac{r}{2\sqrt{\sinh^{2}\left(
r\right)  -r^{2}}}\right]  \text{.} \label{ax2}%
\end{align}
We also obtain the corresponding function $\mathcal{M}$ providing this
generalization for the 't Hooft-Polyakov monopole:
\begin{equation}
\mathcal{M}\left(  r\right)  =\frac{3w_{0}\sinh^{2}\left(  r\right)  }%
{3w_{0}\sinh^{2}\left(  r\right)  -\left(  3w_{0}-1\right)  r^{2}}\text{.}
\label{dsm1}%
\end{equation}
Note that the deformed solutions eqs. (\ref{dsh1}), (\ref{dsw1}) and
(\ref{dsm1}) were already obtained in eqs. (16), (17) and (18) of Ref.
\cite{PLB}. Also, we point out that $w_{0}=1/3$ leads us back to the usual 't
Hooft-Polyakov monopole.

Our second example illustrates the deformation procedure of a completely
analytical scenario involving a second type of monopoles introduced in Ref.
\cite{PLB}, i.e., those ones which can not be reduced to the 't Hooft-Polyakov
solution. For instance, let us consider the self-dual generalized profile%
\begin{equation}
H\left(  r\right)  =\frac{r}{1+r}\text{,} \label{x1}%
\end{equation}
and the corresponding solution for $W\left(  r\right)  $
\begin{equation}
W\left(  r\right)  =\frac{1}{\sqrt{1+r^{2}e^{2r}}}\text{,} \label{x2}%
\end{equation}
whilst the generalization function $M\left(  r\right)  $ is%
\begin{equation}
M\left(  r\right)  =\frac{(1+r)^{2}e^{2r}}{1+r^{2}e^{2r}}\text{.} \label{x3}%
\end{equation}

Now, taking the deformation function%
\begin{equation}
f\left(  \mathcal{H}\right)  =\mathcal{H}^{1/n}(r)\text{,}%
\end{equation}
with real $n>0$, one achieves%
\begin{equation}
\mathcal{H}(r)=\frac{r^{n}}{\left(  1+r\right)  ^{n}}\text{.} \label{H2}%
\end{equation}
Then, from eqs. (\ref{afr}) and (\ref{n}), we obtain%
\begin{equation}
\mathcal{P}\left(  r\right)  =\frac{n\,r^{n+1}}{(1+r)^{n+1}}\text{,}%
\end{equation}%
\begin{equation}
\mathcal{N}\left(  r\right)  =-\frac{\ln(r)}{n}-\frac{r}{n}\text{,}%
\end{equation}
whereas eqs. (\ref{nw}) and (\ref{vd}) yield%
\begin{equation}
\mathcal{W}\left(  r\right)  =\frac{C_{1}}{\sqrt{C_{1}^{2}+r^{2/n}e^{2r/n}}%
}\text{,} \label{W2}%
\end{equation}%
\begin{equation}
\mathcal{M}\left(  r\right)  =\frac{r^{2/n}(1+r)^{n+1}e^{2r/n}}{nr^{n+1}%
(1+r^{2/n}e^{2r/n})}\text{.} \label{M2}%
\end{equation}
Here, $C_{1}$ is a positive real constant. In this case, the family of models
defined by (\ref{M2}) has the analytical self-dual solutions (\ref{H2}) and
(\ref{W2}), which are generalizations of the nonstandard solutions (\ref{x1})
and (\ref{x2}). In particular, for $n=C_{1}=1$, eqs. (\ref{H2}), (\ref{W2})
and (\ref{M2}) reduce to (\ref{x1}), (\ref{x2}) and (\ref{x3}), respectively.

Having analyzed two entirely analytical examples, we now focus our
attention\ on the more sophisticated case\ in which one deforms a partially
analytical configuration into a completely analytical one;\ in this case, only
the Higgs field has a starting analytical profile. We can\ easily verify
whether a particular profile function $H\left(  r\right)  $ gives rise to a
completely analytical configuration or not. The answer is obtained by
combining the BPS equations (\ref{bps1}) and (\ref{bps2})\ into one single
equation, i.e.,
\begin{equation}
\frac{dW}{dr}\frac{dH}{dr}=\frac{\left(  W^{2}-1\right)  HW}{r^{2}}\text{,}
\label{ke}%
\end{equation}
relating $W\left(  r\right)  $ and $H\left(  r\right)  $. Therefore, for a
given $H\left(  r\right)  $, Eq. (\ref{ke}) can be integrated analytically or
not, providing the corresponding solution for $W(r)$ (and vice-versa). We now
consider a case for which Eq. (\ref{ke}) can not be integrated
analytically,\ illustrating the deformation of a model described by the
following analytical expression:
\begin{equation}
H\left(  r\right)  =\frac{\sin\left(  H_{_{tHP}}\right)  }{\sin\left(
1\right)  }\text{,} \label{is}%
\end{equation}
with $H_{_{tHP}}$ being given by (\ref{u1aa}). Note that the denominator
$\sin\left(  1\right)  $\ works as a normalization factor that assures
$H\left(  \infty\right)  =1$. It is worthwhile to point out that, despite the
arbitrariness of the non-Abelian model (\ref{1}), an arbitrary function of
$H_{_{tHP}}$\ is not, in general, a legitimate solution of the generalized model.

The behavior of $W(r)$ around the boundary values (\ref{cc1}) and (\ref{cc2})
can be inferred by using (\ref{ke}). This way, one finds that,\ near the
origin, $W\left(  r\right)  $ can be approximated by
\begin{equation}
W\left(  r\right)  =1-\frac{1}{2}w_{0}r^{2}+\ldots\text{,}%
\end{equation}
whilst, for $r\rightarrow\infty$, it reads%
\begin{equation}
W\left(  r\right)  =\frac{w_{\infty}r}{e^{r\tan\left(  1\right)  }}%
+\ldots\text{,}%
\end{equation}
where $w_{0}$ and $w_{\infty}$ are real constants to be fixed by requiring the
desired behavior near the origin and at infinity, respectively. However, when
we fix $w_{0}$, the parameter $w_{\infty}$\ is automatically fixed, and
vice-versa. Hence, we see that the solutions characterizing the partially
analytical configuration we will deform reach the physical boundary conditions
in the same way (despite numerical factors) as the usual 't Hooft-Polyakov
solution do. In this sense, the topological stability of our initial
configuration is achieved in the standard manner, being verified by the
numerical solution for $W\left(  r\right)  $\ shown in Fig. 2, for $w_{0}%
=1/3$.
\begin{figure}[ptb]
\includegraphics[{width=8.7cm}]{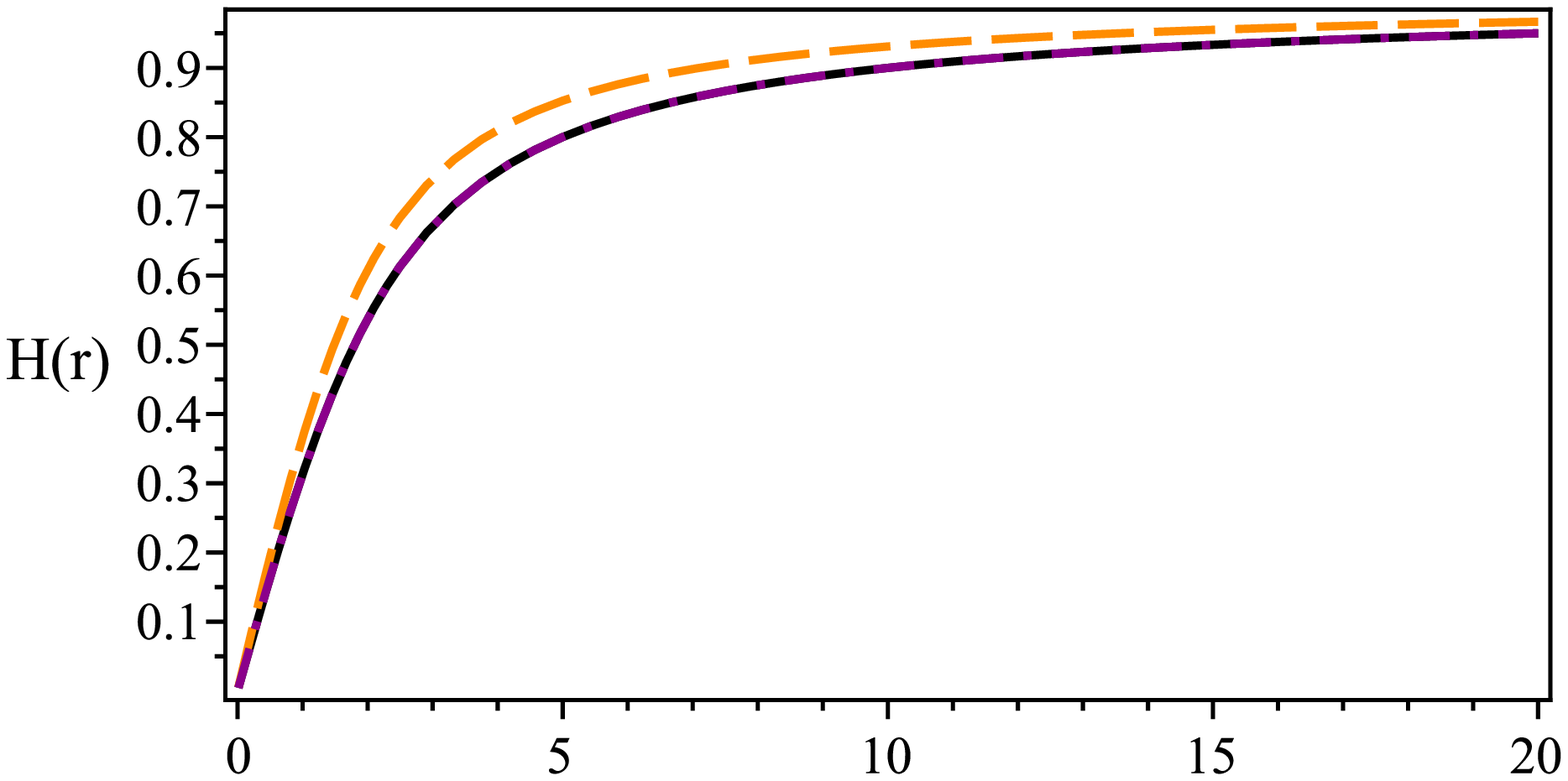}
\includegraphics[{width=8.7cm}]{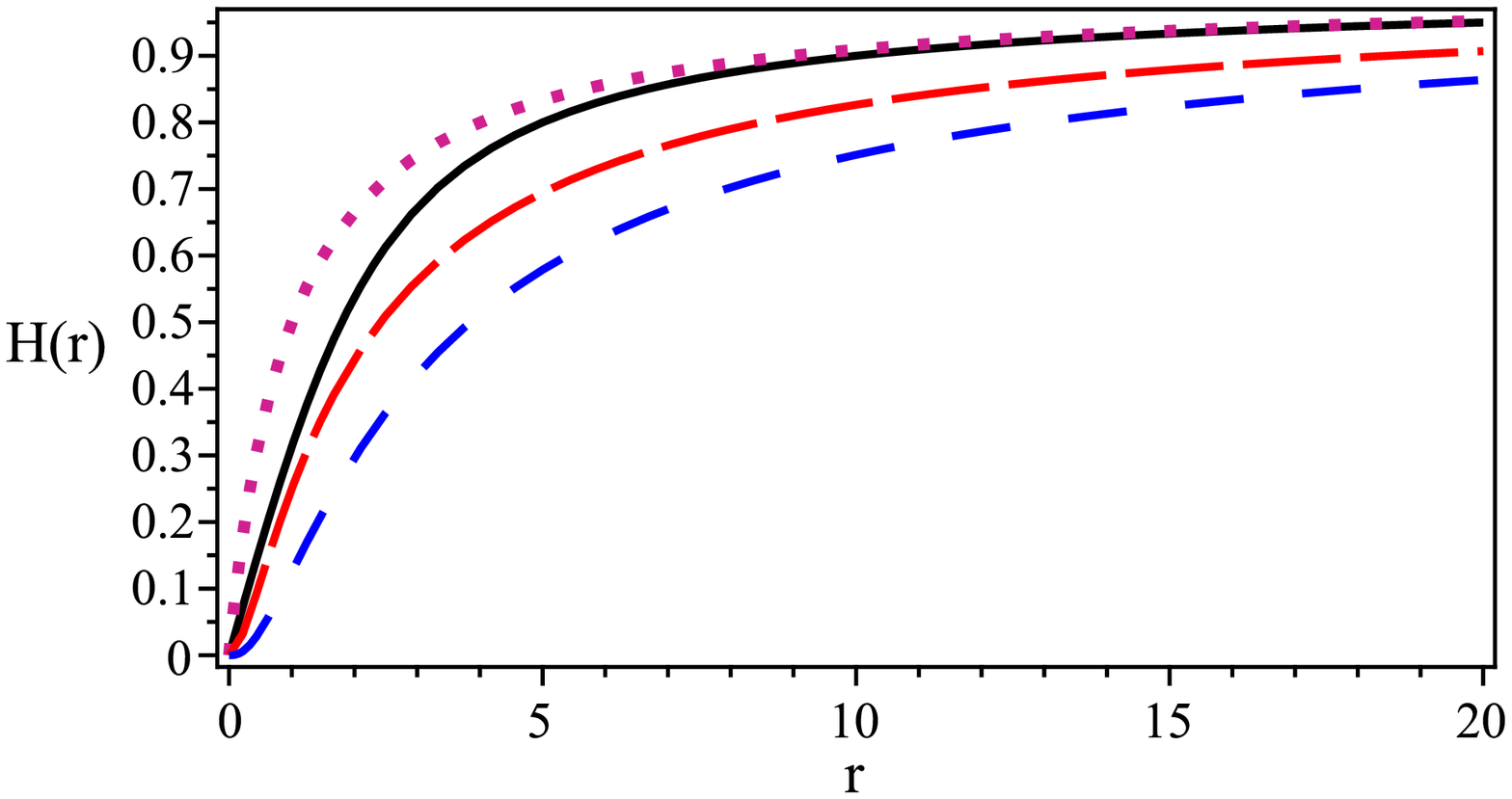}\vspace{-0.25cm}\caption{The Higgs
profile. Top: the solution given by (\ref{is}) (dashed orange line). Bottom:
the solutions given by (\ref{x1}) (dotted purple line) and (\ref{H2})
(long-dashed red line for $n=2$, and space-dashed blue line for $n=3$).\ In
both pictures, the usual profile Eq. (\ref{u1aa}) is also shown (solid black
line), for comparison.}%
\end{figure}

Now, following our prescription, we choose the deformation function as
\begin{equation}
f\left(  \mathcal{H}\right)  =\frac{\sin\left(  \mathcal{H}\left(  r\right)
\right)  }{\sin\left(  1\right)  }\text{.} \label{ip}%
\end{equation}
Then, by combining eqs. (\ref{is}) and (\ref{ip}), we get that the deformed
solution for $\mathcal{H}\left(  r\right)  $ reads as
\begin{equation}
\mathcal{H}\left(  r\right)  =H_{_{tHP}}\left(  r\right)  \text{,}%
\end{equation}
i.e., it\ is the usual 't Hooft-Polyakov solution, whereas the corresponding
deformed solution for $\mathcal{W}\left(  r\right)  $ is%
\begin{equation}
\mathcal{W}\left(  r\right)  =\frac{r}{\sqrt{3w_{0}\sinh^{2}\left(  r\right)
-\left(  3w_{0}-1\right)  r^{2}}}\text{,}%
\end{equation}
exactly the solution previously studied in Eq. (\ref{dsw1}). Hence, the
auxiliary functions $\mathcal{P}\left(  r\right)  $ and $\mathcal{N}\left(
r\right)  $ are given by eqs. (\ref{ax1}) and (\ref{ax2}), respectively.

The overall conclusion is that, by choosing suitable deformation functions,
both completely and partially analytical monopole configurations can be
deformed into a new analytical one.

In the sequel, we depict all the solutions we have found and compare them with
the usual 't Hooft-Polyakov one, commenting on the main features of the new
profiles. The solutions for the Higgs field are shown in Fig. 1, which reveals
that all the profiles exhibit the same general behavior, reaching their
boundary values monotonically, whilst spreading over different distances.
\begin{figure}[ptb]
\includegraphics[{angle=0,width=8.7cm}]{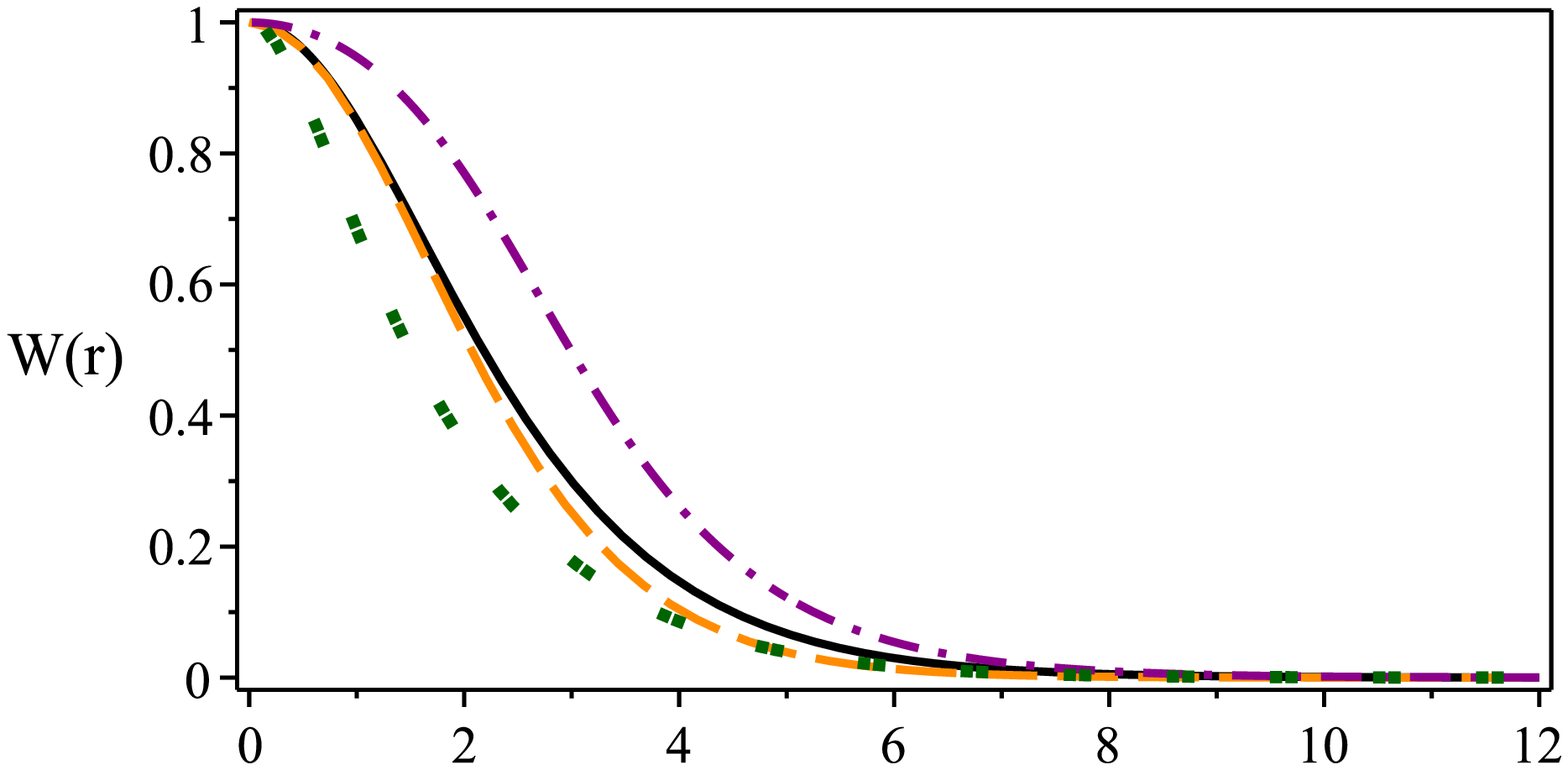}
\includegraphics[{angle=0,width=8.7cm}]{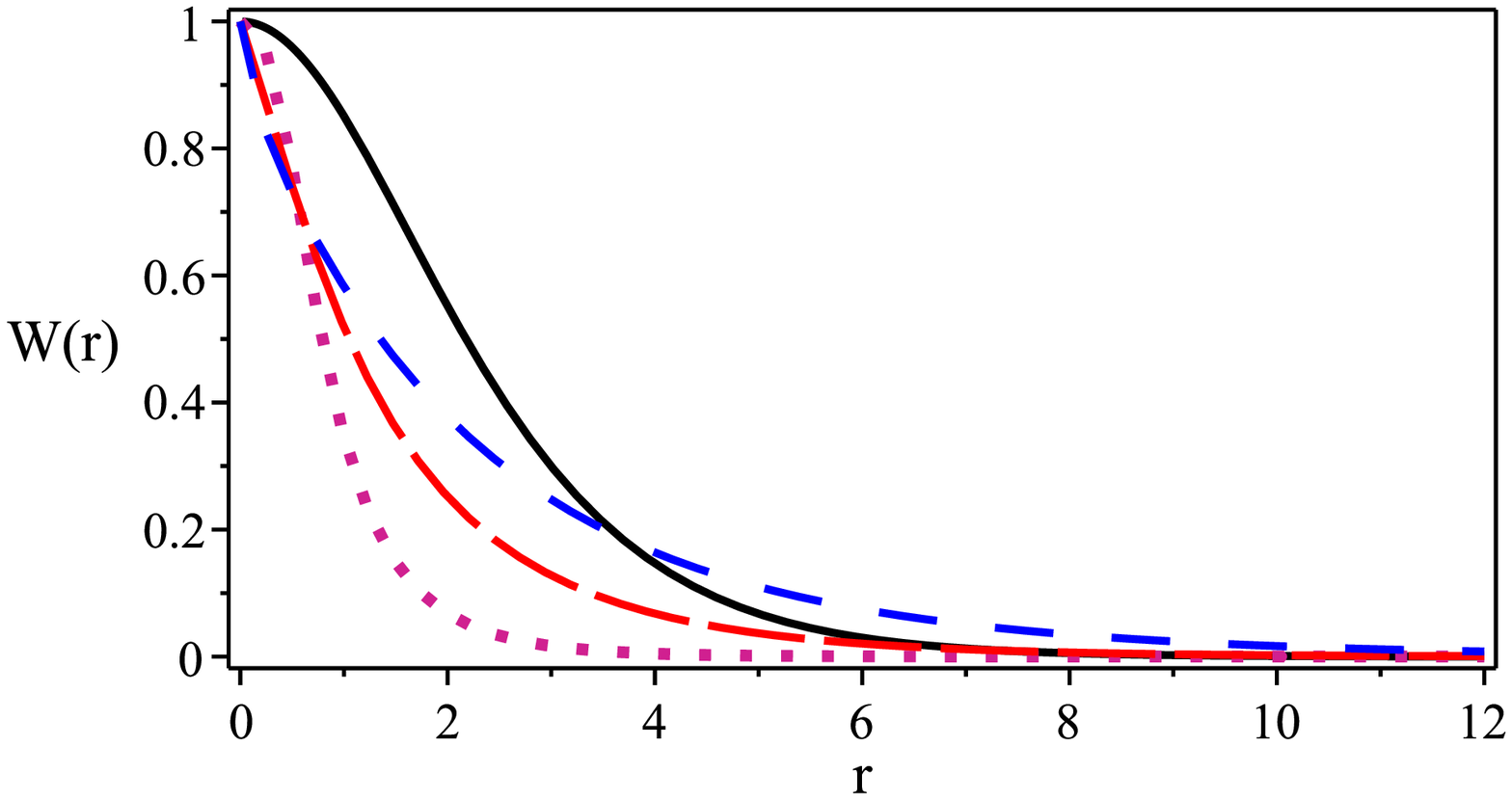}\vspace{-0.25cm}%
\caption{The gauge profile. Top: the solutions given by Eq. (\ref{dsw1}) for
$w_{0}=1/10$ (dot-dashed dark-purple line), and $w_{0}=1$ (double-dotted green
line). The numerical profile related to the Eq. (\ref{is}) is also shown
(dashed orange line for $w_{0}=1/3$). Bottom: the profiles given by (\ref{x2})
and (\ref{W2}). Here, the conventions are the same as in Fig. 1, with
$C_{1}=1$.}%
\end{figure}
In Fig. 2, we depict the solutions for the gauge field. Here, by plotting Eq.
(\ref{dsw1}), we identify the way the integration constant coming from
(\ref{nw}) controls the characteristic length of the deformed solutions: the
profiles for $w_{0}>1/3$ spread over smaller distances, while exhibiting
greater cores for $0<w_{0}<1/3$.
\begin{figure}[ptb]
\includegraphics[{angle=0,width=8.7cm}]{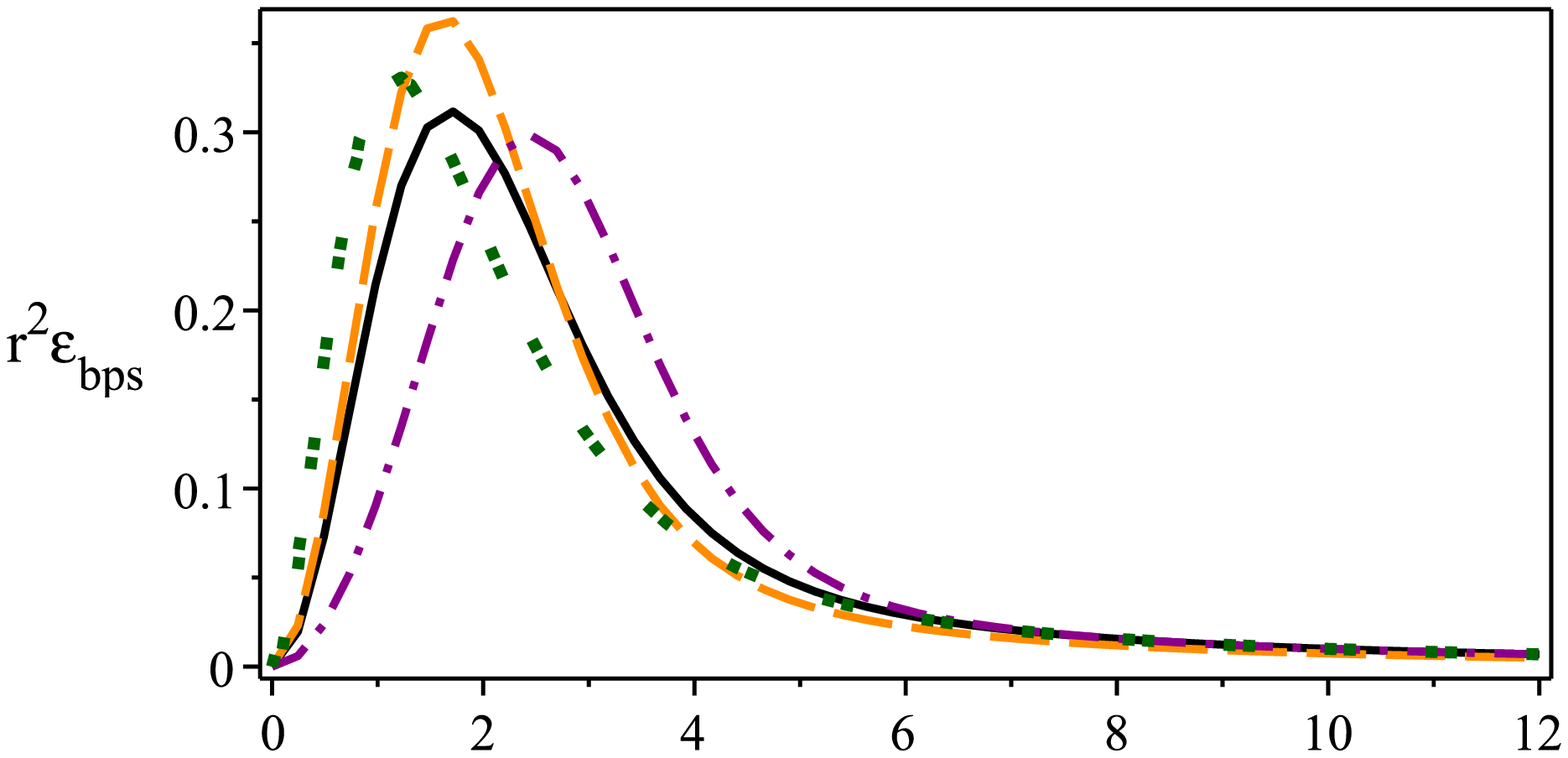}
\includegraphics[{angle=0,width=8.7cm}]{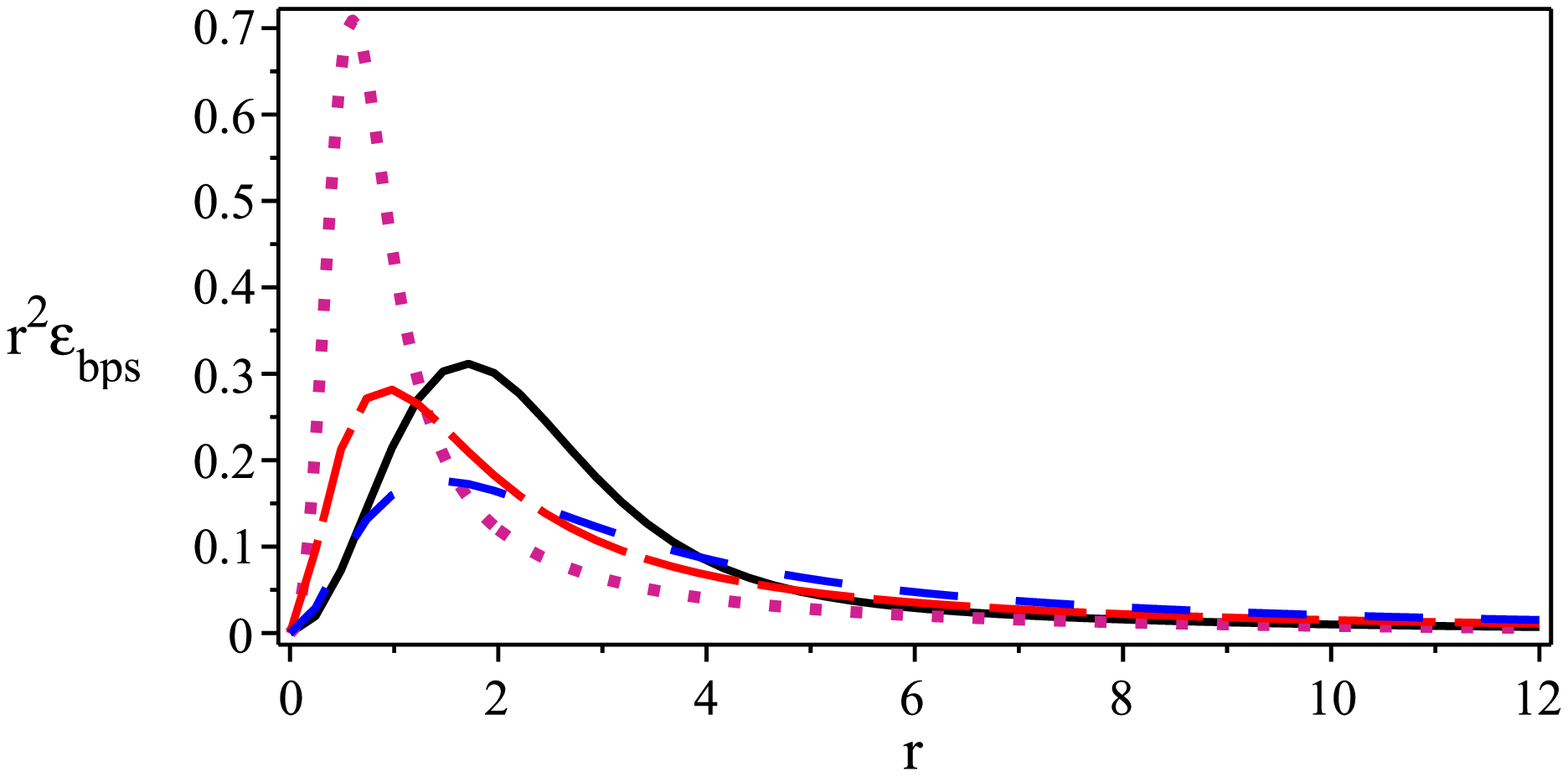}\vspace{-0.25cm}%
\caption{The product $r^{2}$ times the BPS energy density. Conventions as in
Fig. 2.}%
\end{figure}

Fig. 3 displays the results for the product $r^{2}$ times the BPS\ energy
densities. These profiles are important, since the enclosed area gives (within
a constant factor of $4\pi$) the energy of the BPS monopoles. In general, the
solutions we have found are rings, i.e., they reach the corresponding
amplitudes at some finite distance $R$ from the origin.

Finally, we point out that the new solutions we have found, despite
well-behaved, are of the type $H\left(  r\right)  $ or $H\left(  r\right)  .$
To obtain the inverse function, $r(H)$ or $r(H),$ provides fairly complicated
relations. This way, we have preferred to express $M$\ or $M$ as explicit
functions of $r$. This fact does not prevent the existence of simpler
configurations for $r(H)$, which were still not found out, however.


\section{Ending comments}

\label{end}

In this work, we have established a deformation prescription consistent with
the generalized self-dual Yang-Mills-Higgs scenario presented by some of us in
a recent paper \cite{pau}. Here, starting from well-known BPS field profiles,
the deformation procedure allows to obtain new self-dual solutions standing
for the magnetic monopoles arising within a non-Abelian-Higgs model endowed by
a particular positive function $M$. It is worthwhile to point out that the
initial configuration can be completely or partially analytical, the final
scenario possessing exact solutions for both gauge and scalar fields.

We have checked our algorithm by studying some illustrative examples. The
first two cases we have considered were entirely analytical, one based on the
usual 't Hooft-Polyakov solution, the other based on the nontrivial solution
introduced in \cite{PLB}. In the sequel, we have extended our work for a
partially analytical configuration. It is important to point out that we have
implemented a deformation prescription which gives legitimate new self-dual
solutions of a different model with similar BPS equations. Such deformed
solutions can not be attained by a trivial redefinition of the standard fields.

The results we have found are depicted in figs. 1, 2, and 3, the overall
conclusion being that the deformed profiles behave in the same general way
their standard counterpart do. In figs. 2 and 3, we have also identified the
way the constant of integration $w_{0}$ affects the resulting profiles.

We are now investigating the possibility to develop a deformation procedure
applicable to the study of self-dual Maxwell-Higgs, Chern-Simons-Higgs and
Maxwell-Chern-Simons-Higgs vortices. We hope to report on this in the near future.

The authors thank CAPES, CNPq and FAPEMA (Brazilian agencies) for partial
financial support.

\end{document}